\documentstyle[prl,aps,twocolumn,EPSF]{revtex}
\newcommand{\be}{\begin{equation}}
\newcommand{\ee}{\end{equation}}
\newcommand{\bea}{\begin{eqnarray}}
\newcommand{\eea}{\end{eqnarray}}

\begin{document}
\draft
\title{Large moments in external fields of cubic group symmetry}
\author{V.~A.~Kalatsky$^a$ and V.~L.~Pokrovsky$^{a,b}$}

\address{
$^a$ Department of Physics, Texas A\& M University, 
College Station, Texas 77843-4242
}
\address{
$^b$ Landau Institute for Theoretical Physics, 
Kosygin str.2, Moscow 117940, Russia
}
\date{\today}
\maketitle
%\narrowtext
\begin{abstract}
%The study of large spins $S$, in strongly anisotropic external fields 
%of point group symmetry ${\bf O}$ (octahedron), 
%reveals a dissimilarity of the spin 
%behavior inside of the two statistical classes, {\em e.g.}, properties of 
%the odd-integer spins are different from those of the even-integer. 
%We propose a scenario of the experiment where the distinction can be found. 

We predict that large moments $J$, placed into a crystal field with 
the cubic point symmetry group, differ by their spectrum and 
magnetic properties. 
{\em E. g.}, properties of the odd-integer moments are different 
from those of the even-integer. 
The effect is due to Berry's phases gained by the moment, 
when it tunnels between minima of the external field. 
Two cases of the group ${\bf O}$ are classified, 
namely, 6- and 8-fold coordinations. 
The spectrum and degeneration of energy levels depend 
on a remainder $\{J/n\}$, where 
the divisor $n=4$ and 3 for 6-fold and 8-fold coordination respectively.
%High symmetry results in a finite magnetic moment for half-integer 
%and some integer moments, for example odd $J$ at 6-fold coordination. 
Large moments in the cubic environment can be realized by diluted alloys 
$\mbox{R}_{1-x}\mbox{R}_{x}'$Sb, where R=Lu, La, and 
R$'$=Tb, Dy, Ho, Er.

\end{abstract}

%\section{Introduction}
It has been shown on several examples (dependence of the Haldane gap in 
anti-ferromagnetic spin chain~\cite{Haldane} and in 2D isotropic 
Heisenberg antiferromagnets~\cite{Haldane1}, suppression of tunneling 
in magnetic particles~\cite{Loss}, magnetization of ions 
in external fields~\cite{local1}) 
that systems with large integer moments may behave differently from 
ones with large half-integer moments although the values of 
the moments are close to each other. 
Haldane pointed out in~\cite{Haldane1} that there may be a 
difference between odd and even spins as well.
All of them are due to Berry's connection~\cite{Berry}. 
In this paper we demonstrate 
systems where further differentiation on the moment values occurs. 

The systems under study 
are large moments in strongly anisotropic external fields, 
{\em e.g}, crystal electric field with deep wells at certain crystallographic 
directions. We consider systems of the octahedron symmetry with 6- and 
8-fold coordination. 
They show a subtle partition on the moment values. 

%\section{6-fold or octahedral coordination}

\begin{figure}
  \centerline{\epsfxsize=1.5in \epsfysize=1.5in 
\epsffile{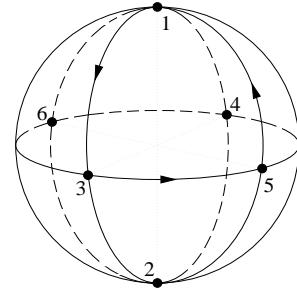}}
  \caption{ Paths of the spin on the unit sphere between 
the easy positions of the field. The case of the 6-fold coordination.}
  \label{fig:1}
\end{figure}

In classical description, the angular moment is a vector of a fixed length 
$J$ determined by the direction of a unit vector ${\bf n}$.
Let us assume that an ion with a large moment is placed into 
a crystalline electric field (CEF). Due to symmetry  
the moment may be localized in 6 symmetrical positions 
(the easy positions of CEF). We choose following spherical coordinates 
($\theta$, $\phi$) for the positions: 1-(0, 0), 2-($\pi$, 0), 
3-($\pi/2$, 0), 4-($\pi/2$,$\pi$), 5-($\pi/2$, $\pi/2$), 
and 6-($\pi/2$, $3\pi/2$). 
Such a configuration will be called octahedral or 6-fold. 
The ground state of the system is 6-fold degenerate if tunneling between 
the potential wells is neglected. 
If one allows the moment to tunnel to the nearest wells along the geodesics 
on the unit sphere (see Fig.~\ref{fig:1}), 
the degeneracy will be lifted at least partially. 
A formal description of such a system is given by a 
Hamiltonian (${\cal H}$) with the following matrix elements:
\bea
h_{ii}&=&0,\,\,i=1,\,2,\ldots 6, \nonumber \\  \label{Ham_6}
h_{ij}&=&0,\,\,|i-j|=1,\,i+j=3,\,7,\,11, \\
|h_{ij}|&=&|w|,\,\,\mbox{for other } 1\leq i,j\leq6\nonumber
\eea
where we adopted the enumeration shown on Fig.~\ref{fig:1}. 
 
The moment's wave-function, upon tunneling from one minimum 
($i$) to another ($j$), 
may gain a geometric phase shift $\phi(C_{ij})$ 
which depends on the tunneling path $C_{ij}$ on the unit sphere 
(the ${\bf n}$-space): 
$\phi(C_{ij})=\int_{C_{ij}}{\bf A}({\bf n})\mbox{d}{\bf n}$. 
The vector field ${\bf A}$ is called Berry's connection. Specially 
for the problem of semi-classical moments it was calculated 
in~\cite{Berry,local1}.
The appropriate geometric phase was first discovered by Berry~\cite{Berry}. 

The phase shifts $\phi(C_{ij})$ as well as the connection 
${\bf A}$ are not gauge invariant:
${\bf A}({\bf n})\rightarrow{\bf A}({\bf n})-\nabla f({\bf n})$,
where $f({\bf n})$ is an arbitrary continuous function, 
the spins' eigenfunctions transform as:
$\psi({\bf n})\rightarrow e^{if({\bf n})}\psi({\bf n})$. 
In our case (well localized states) it reduces to: 
$\psi_k\rightarrow e^{if_k}\psi_k$, 
where $f_k=f({\bf n}_k)$, ${\bf n}_k$ ($k=1,\ldots,6$) are the positions of 
the potential wells.
However, the phase $\gamma(C)$ accumulating at circulation along a 
closed circuit $C$ is invariant: 
\be
\gamma(C)=\oint_C{\bf A}({\bf n})\cdot\mbox{d}{\bf n}=\pm J\Omega(C),
\label{phase}
\ee
where $\Omega(C)$ is the solid angle subtended by a closed 
path $C$ on the unit sphere at its origin. 
The phase shifts $\phi(C_{ij})$ change to $\phi(C_{ij})+f_j-f_i$ 
by the gauge transformation. Therefore, $\phi(C_{ij})$ can be 
any set of numbers satisfying the invariant relationships~(\ref{phase}). 
The Hamiltonian also is not invariant, it transforms as: 
${\cal H}\rightarrow{\cal UHU^{\dagger}}$, 
where ${\cal U}$ is a diagonal matrix with matrix elements
$u_{kk}=\exp(if_k)$.
 
%The only condition imposed on the phase shifts is that their oriented 
%sum over a closed path should be equal to $J$ times solid angle, 
%that the path subtends at the origin, modulo $4\pi$. 

%We shall call a closed path covering minimal non-zero solid angle 
%a basic loop. A closed path 1-3-5-1 on Fig.~\ref{fig:1} is 
%an example of a basic loop. Due to the symmetry of our problem, 
%the parameter space can be covered by 8 basic loops. 
%Each one subtends the solid angle of $\pi/2$. 
Tunneling trajectories divide the sphere into 8 plaquettes. 
Each of them represents a minimal non-trivial closed loop and 
subtends the solid angle of $\pi/2$ 
({\em e. g.} closed path 1-3-5-1 on Fig.~\ref{fig:1}).
The 8 plaquettes give 8 equations for the phase shifts:
\be
\phi_{ij}+\phi_{jk}+\phi_{ki}=J(\frac{\pi}{2}\,\mbox{mod}\,4\pi),
\label{phase-shifts6}
\ee
where $\phi_{ij}$ is the phase shift upon 
tunneling from easy position $i$ to easy position $j$, and 
set $\{(i,j,k)\}$ enumerates the plaquettes.
%$=(1,3,5)$, $(1,5,4)$, $(1,4,6)$, $(1,6,3)$, $(2,5,3)$, 
%$(2,4,5)$, $(2,6,4)$, $(2,3,6)$. 
Only 7 out of the 8 equations 
are independent which leads to 5 degrees of freedom over 12 phases. 
This number (five) is equal to the number of independent parameters 
in a general gauge transformation (six) minus one corresponding 
to a common phase factor.
%These 5 degrees of freedom are embedded in the gauge 
%transformation ${\cal U}$. To show it one can rewrite ${\cal U}$ as 
%$\exp(if_1){\cal U'}$ where ${\cal U'}$ depends on the 5 differences 
%$f_k-f_1,\,k=2,\ldots,6$, the common factor $\exp(if_1)$ does not change 
%the Hamiltonian.
%To fix the gauge we chose the following rather symmetric set of phases: 
%\bea
%\phi_{13}=\phi_{35}=\phi_{51}=-\frac{7\pi}{6},\,\, 
%\phi_{26}=\phi_{64}=\phi_{42}=\frac{\pi}{6}, \nonumber \\
%\phi_{14}=\phi_{52}=\phi_{36}=-\frac{2\pi}{3},\,\,
%\phi_{23}=\phi_{61}=\phi_{45}=\frac{4\pi}{3}.
%\label{gauge} 
%\eea
%In this choice of the gauge the Hamiltonian of the 6-fold system is
%\be
%w\left|
%\begin{array}{cccccc}
%0&0&e^{iJ\phi_{13}}&e^{iJ\phi_{14}}&e^{-iJ\phi_{51}}&e^{-iJ\phi_{61}} \\
%0&0&e^{iJ\phi_{23}}&e^{-iJ\phi_{42}}&e^{-iJ\phi_{52}}&e^{iJ\phi_{26}} \\
%e^{-iJ\phi_{13}}&e^{-iJ\phi_{23}}&0&0&e^{iJ\phi_{35}}&e^{iJ\phi_{36}} \\
%e^{-iJ\phi_{14}}&e^{iJ\phi_{42}}&0&0&e^{iJ\phi_{45}}&e^{-iJ\phi_{64}} \\
%e^{iJ\phi_{51}}&e^{iJ\phi_{52}}&e^{-iJ\phi_{35}}&e^{-iJ\phi_{45}}&0&0 \\
%e^{iJ\phi_{61}}&e^{-iJ\phi_{26}}&e^{-iJ\phi_{36}}&e^{iJ\phi_{64}}&0&0 
%\end{array}
%\right|,
%\label{Ham6}
%\ee
%A remark on the rotational invariance: 
Note that the Hamiltonian~(\ref{Ham_6}) is 
not invariant under the action of the octahedron group 
realized as a subgroup of the permutation group on the 6 states. 
What leaves the Hamiltonian 
invariant is a rotation (${\cal R}$) accompanied by an appropriate 
gauge transformation (${\cal U}$):
${\cal H}={\cal U}{\cal R}{\cal H}{\cal R}^{T}{\cal U}^{\dagger}$,
where ${\cal U}$ depends on ${\cal R}$, and $J$. 

Hamiltonian~(\ref{Ham_6}) can be diagonalized leading 
to the following 6 eigenvalues:
\bea
\label{exact6}
&E_k&(J)=(-1)^k2w\chi(\pi(J+2k)),\,\,k=0,\ldots,5, \\
&\chi&(x)=\cos\frac{2x}{3}\cos\frac{x}{2}-
\left(\cos^2\frac{x}{3}+
\sin^2\frac{2x}{3}\sin^2\frac{x}{2}\right)^{1/2}. \nonumber
\eea
The ground state energies and their degeneracies fall into 5 classes 
which are gathered in Table~\ref{tab:1}.
The classes differ by their ground-state energies, degeneracies of 
the ground-state, and excitation energies. All these will lead to 
experimentally observable dissimilarities  between systems falling 
into the different classes for $k_{\rm B}T<w$, {\em e. g}, 
thermodynamical properties or response to applied magnetic field. 

The number of the classes and the moments' values of each class can be 
easily found without detailed analysis of the Hamiltonian for the 
external fields of the perfect polyhedra group symmetries 
({\em perfect Hamiltonians}). 
Let us assume that the parameter space (unit sphere) can be covered 
with $p$ plaquettes (the number of the faces of a perfect polyhedron). 
Then, each plaquet subtends a solid angle of 
$4\pi/p$ at the origin of the sphere and Berry's phase for each 
loop is $4\pi J/p$. This gives a period of $J$: $J_p=p/2$. 
The spectra of the systems differing by transformation 
$\gamma(C)\rightarrow -\gamma(C)$ must be identical due to 
time-reversal symmetry. 
It leads to equivalence of $J=J_pn+m/2$ and $J=|J_pn-m/2|$, 
$n,\,\,m=0,1,2,\ldots$. 
Hence, we can classify equivalent moments in the following manner:
$$
J_m=|J_pn\pm \frac{m}{2}|,\,\,n=0,\,1,\,2,\ldots;\,\,m=0,\,1,\,\ldots,J_p,
$$
and the number of the classes is $C_p\equiv J_p+1=p/2+1$. 
The classes can be conveniently labeled by the integer $m+1$.
{\em E. g.}, for the 6-fold coordination (octahedron) one has: 
$p=8$, $C_8=5$. 
Another example is a planar external field 
(degenerate polyhedron with two faces - polygon): 
$p=2$, $C_2=2$, and $J_m=|n\pm m/2|$, $m=0,\,1$. This case was studied 
in~\cite{local1} for polygon=square and in~\cite{Loss} for 
polygon=line (easy plane, easy axis anisotropy).

There is an extra symmetry in systems with a perfect Hamiltonian 
which may relate energy levels of different classes or inside of a class. 
It is the change of sign of $w$. This transformation should inverse energy 
levels inside of each class. On the other hand, all physical quantities depend 
only upon gauge invariant combinations of a type $w^k\cos(J\Omega_k)$, 
where $\Omega_k$ is the solid angle subtended by a 
closed loop of $k>1$ walks along the geodesics. In some cases 
(polyhedra made of even-sided faces: cube, square) 
$k$ is always even and this leads to the symmetry of the levels inside 
of each class, namely, the levels must come in pairs of opposite sign $\pm E$. 
In other occurrences (polyhedra made of odd-sided faces: 
octahedron, tetrahedron) $k$ may be both odd and even. 
%However, in the latter systems one always can chose such a gauge that all 
%$\phi_{ij}$ are odd multiples of $4\pi/p$. 
Then simultaneous change 
of sign of $w\rightarrow -w$ and shift of $J\rightarrow J+p/4$ 
leaves the invariant combinations unchanged. 
Therefore, each level $E$ in the class of $J$ has its 
counterpart  $-E$ in the class of $J+p/4$. 
If $J$ and $J+p/4$ are equivalent, 
their spectrum is symmetric ({\em e.g.}, class 3 in Table~\ref{tab:1}).

Let us consider the magnetic response of different spins.
%among the classes we introduce 
%magnetic field (${\bf H}=(H_x\,H_y,\,H_z)$) into the system. 
We assume that magnetic field ${\bf H}$ is small, {\em i.e.}, 
$gJ\mu_B|{\bf H}|\ll U(J)$ where $U(J)$ is the variation 
of the CEF potential. 
%at its minima (due to symmetry the depths are the same).
The magnetic part of the Hamiltonian is a diagonal matrix: 
$$
{\cal H}_{\rm M}=
J\,diag(-h_z,\,h_z,-h_x,\,h_x,-h_y,\,h_y),\,\,h_{\alpha}=g\mu_BH_{\alpha},
%\label{Zeeman}
$$ 
where $g$ is the gyro-magnetic ratio and $\alpha=x,\,y,\,z$. 
The full Hamiltonian (${\cal H}+{\cal H}_{M}$) can be diagonalized 
for some symmetric directions of the field, {\em e. g.}, along an easy 
direction $(0,0,1)$. 
%However, we will not 
%give the explicit formulae for the ground states due to their lengthiness. 
We present the ground state energies up to the second order 
in magnetic field for all classes in Table~\ref{tab:2}.

Note that class 2 has anisotropic magnetic moment, and classes 1, 2, and 3 
have anisotropic magnetic susceptibility in the ground state.
Some classes (2, 3, and 4) have non-zero magnetization 
in the ground-state. The low temperature behavior of these classes 
should be strikingly different from those without magnetization. 
We will discuss it later. 
%after considering the 8-fold coordination system. 
%The degeneracy of the ground state is lifted in the magnetic field 
%for all classes except the first class, where the ground state 
%is twofold degenerate for $(\pm1,\pm1,\pm1)$ directions of magnetic field.

%\section{8-fold or cubic coordination}
 
\begin{figure}
  \centerline{\epsfxsize=1.5in \epsfysize=1.5in 
\epsffile{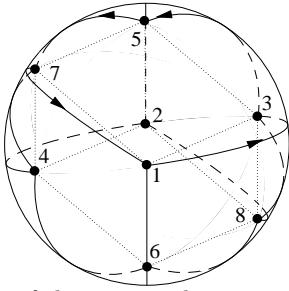}}
  \caption{ Paths of the spin on the unit sphere between 
the easy positions of the field. The case of the 8-fold coordination.}
  \label{fig:2}
\end{figure}

For the 8-fold or cubic case the easy positions of CEF, 
on the unit sphere, can be chosen in the following way: 
$(\pm\arccos(1/\sqrt3),(2k+1)\pi/4),\,\,k=0,1,2,3$ and enumerated as 
$(4k+3\pm1)/2$ (see Fig.~\ref{fig:2}). 
The parameter space can be covered with 
$p=6$ plaquettes and we should expect $C_6=4$ classes with members 
$J_m=|3n\pm m/2|$, $m=0,\,1,\,2,\,3$ all sets of levels are symmetric. 
The solid angle subtended by each plaquet is $2\pi/3$ 
which gives 6 (5 independent) equations for the phase shifts.
%\be
%\phi_{ij}+\phi_{jk}+\phi_{kl}+\phi_{li}=\frac{2\pi}{3}\mbox{mod}4\pi,
%\label{phase-shifts8}
%\ee
%where quadruplets (i,j,k,l) parameterize the basic loops.

The eigenvalues of the Hamiltonian can be expressed in the 
following closed form:
\bea
\label{8exact}
E_k^{\pm}&=&\pm2w\xi(\pi(J+3k)),\,\,k=0,\,1,\,2,\,3, \\
\xi(x)&=&(3+2\cos x\cos\frac{2x}{3}+
4\cos\frac{x}{2}\cos\frac{x}{3}f(x))^{\frac12} \nonumber \\
f(x)&=&(4\sin^2\frac{x}{2}\sin^2\frac{x}{3}+1)^{\frac12}. \nonumber
\eea
The ground state energies of the cubic coordination classes 
and their degeneracies are presented in Table~\ref{tab:3}. 
The responses of different classes to applied magnetic field should 
be quite different.
To demonstrate this we apply magnetic field along one of the 
easy direction of the CEF, {\em e. g.}, the first position on 
Fig.~\ref{fig:2} (${\bf H}=H(1,1,1)/\sqrt3$). 
The Zeeman term of the Hamiltonian, in the chosen 
parameterization, is $-hJ(3,-3,1,-1,-1,1,1,-1)/3$.
Taylor expansions of the ground state energies up to the second 
order in magnetic field are given in Table~\ref{tab:4}.
There are three classes with non-zero magnetization in the 
ground state.
Class 4 has anisotropic magnetic moment, and class 3 
has anisotropic magnetic susceptibility in the ground state.
%The degeneracy of the ground state is lifted in the magnetic field 
%for all classes except the fourth class, where the ground state 
%is twofold degenerate for $(1,0,0)$, and its equivalent, 
%directions of magnetic field.

There exists one more type of the coordination in the cubic CEF --- 12-fold. 
The 12-fold coordination is realized if the minima of the CEF are located 
at the centers of the cube's edges and can be parameterized 
on the unit sphere as 
$(k\pi/4,(2k+l-1)\pi/4)$, $k=1,\,2,\,3$, $l=0,\,1,\,2,\,3$. 
Unfortunately, the tunneling trajectories do not coincide with the geodesics. 
As a consequence, the number of classes, in general, is infinite. 
We will not consider the 12-fold coordination here due to its strong 
dependence on the CEF parameters.

\begin{figure}
  \centerline{\epsfxsize=3.375in \epsfysize=2.0in 
\epsffile{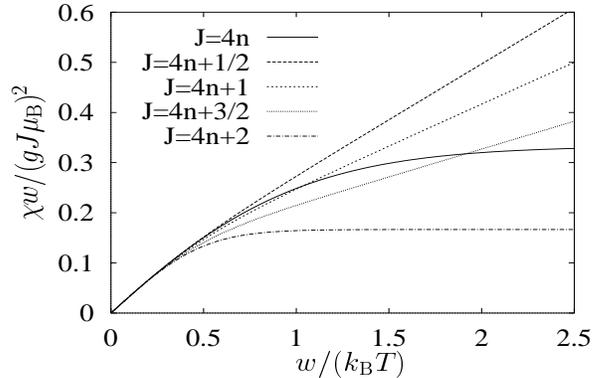}}
\caption{Susceptibility versus inverse temperature in the 6-fold coordination}
%  \caption{ Dimensionless magnetization ($M/(gJ\mu_B)$) vs 
%dimensionless magnetic field ($gJ\mu_BH/w$) ${\bf H}=H(0,0,1)$ 
%for the 5 classes.}
  \label{fig:3}
\end{figure}

\begin{figure}
  \centerline{\epsfxsize=3.375in \epsfysize=2.0in 
\epsffile{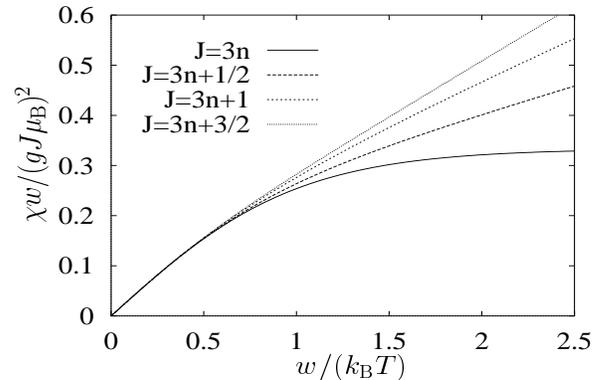}}
\caption{Susceptibility versus inverse temperature in the 8-fold coordination}
%  \caption{ Dimensionless magnetization ($M/(gJ\mu_B)$) vs 
%dimensionless magnetic field ($gJ\mu_BH/w$) ${\bf H}=H(0,0,1)$ 
%for the 5 classes.}
  \label{fig:4}
\end{figure}

Now we proceed to the temperature variation of the magnetic response. 
For temperatures $k_{\rm B}T>w$ large moments become purely classical 
and exhibit Curie magnetic susceptibility: 
\be
\chi_{\rm C}=(\mu_{\rm B}gJ)^2/(3k_{\rm B}T).
\label{Curie}
\ee
For temperatures $k_{\rm B}T<w$ the quantum effects change 
the response drastically. The susceptibility saturates 
for the classes without magnetic moment in the ground state to the value:
\be
\chi_s=\frac{1}{g_0}\sum_{i=1}^{g_0}\chi_{i},
\label{suscept_zero}
\ee
where $g_0$ is the degeneracy of the ground state and 
$\chi_{i}$ is the susceptibility of the $i$th member of the 
ground-state multiplet. 
%Whereas, it has the Curie-like behavior 
For the classes with the magnetic moment in the ground state 
Curie susceptibility persists at $k_{\rm B}T<w$, but its slope 
is different:
\be
\chi_l=\frac{1}{g_0}\sum_{i=1}^{g_0}m_{i}^2/(k_{\rm B}T),
\label{suscept}
\ee
where $m_{i}$ is the moment of the $i$th member of the 
ground-state multiplet. 
We collect the low temperature susceptibilities of the 6- and 
8-fold systems in the last column of 
Tables~\ref{tab:2} and~\ref{tab:4} respectively 
(a common factor is $(gJ\mu_{\rm B})^2$). 
The exact dependencies of the susceptibility versus inverse temperature are 
given on Fig.~\ref{fig:3} and~\ref{fig:4} for the 6- and 8-fold 
coordinations respectively. The susceptibility is isotropic in all 
cases due to the cubic symmetry.

Now, let us deliberate experimentally observable 
differences among the classes. 
First of all we need to have a rare-earth compound with 
the octahedral coordination.
The general form of the CEF potential of the octahedral coordination, 
as it was shown in~\cite{CEF}, reads:
\bea 
U(J_x,J_y,J_z)=A(J_x^4+J_y^4+J_z^4-\frac35J^4)+ \nonumber \\
B(J_x^6+J_y^6+J_z^6+30J_x^2J_y^2J_z^2-\frac57J^6)/J^2,
\label{CEF1}
\eea
where $A$ and $B$ are CEF constants. 
Higher order invariants are irrelevant for the most interesting 
situation of the rare-earth (R) ions with $4f$-electrons.

A general analysis of the CEF potential~(\ref{CEF1}) shows that 
the plane of parameters ($A$-$B$) is divided into three sectors by 
the regions of stability of the 6-, 8-, and 12-fold 
coordinations:
\bea
\mbox{6-fold ($\Gamma_c$): }  & A<B/3,\,\,A<-3B/2, \nonumber \\
\mbox{8-fold ($\Gamma_c^v$): }  & A>B/3,\,\,A>35B/6, \label{sectors} \\
\mbox{12-fold ($\Gamma_c^f$): } & -3B/2<A<35B/6. \nonumber
\eea
where we enclose in parentheses the Bravais lattices of the cubic system 
which realize the respective coordinations. 
An example of the 6-fold coordination is RSb rock-salt 
structured family of compounds. 

The experimental observation of the discussed effects can be 
done on dilute alloys of the type R$_x$R$_{1-x}'$Sb, where 
R$'$=La, Lu (non-magnetic), and 
R=Tb ($J=6$, class 5), Dy ($J=15/2$, class 2), Ho ($J=8$, class 1) 
or any other equivalent set of elements given in Table~\ref{tab:5}. 
Besides magnetization and susceptibility measurements discussed above 
one can study the EPR in these alloys. The spectral measurements can 
display the universal ratios of the resonance frequencies and 
check the values of the magnetic momenta predicted by the theory.
Our preliminary estimates show that Jahn-Teller effect is not 
essential in the conditions of the experiment proposed in this 
article at least at temperatures higher than 0.1K. 
The analysis of Jahn-Teller effect will be done elsewhere.

This work was supported by the NSF grant DMR-9705812. 
We thank P.C.~Canfield who indicated us RSb compounds' family.

\begin{table}
\caption{The ground state energies and their degeneracies of the 
6-fold coordination for $w>0$ ($n=1,\,2,\,3,\ldots$)}
\begin{tabular}{lllll}
Class&Moment ($J$)&\multicolumn{2}{c}{Ground state}&First excited\\ 
     &      &energy&degeneracy&state energy       \\\hline
1 & $4n$         & $-2w$         & 2 & $0$        \\
2 & $4n\pm1/2$   & $-\sqrt2w$    & 4 & $2\sqrt2w$ \\
3 & $4n\pm1$     & $-2w$         & 3 & $2w$       \\
4 & $4n\pm3/2$   & $-2\sqrt2w$   & 2 & $\sqrt2w$  \\
5 & $4n+2$       & $-4w$         & 1 & $0$        \\
\end{tabular}
\label{tab:1}
\end{table}

\begin{table}
\caption{The ground state energies in magnetic field along 
$(0,0,1)$ direction and the low temperature magnetic susceptibilities 
of the 6-fold coordination classes ($\beta=1/(k_{\rm B}T)$)}
\begin{tabular}{lll}
Class&Ground state energy&Susceptibility\\ \hline
1 & $-2w-J^2h^2/(3w)$ 			& $1/(3w)$\\
2 & $-\sqrt2w-2Jh/3-\sqrt2J^2h^2/(27w)$ & $2\beta/9$\\
3 & $-2w-Jh/2-J^2h^2/(16w)$ 		& $\beta/6$ \\
4 & $-2\sqrt2w-Jh/3-\sqrt2J^2h^2/(27w)$ & $\beta/9$ \\
5 & $-4w-J^2h^2/(12w)$ 			& $1/(6w)$  \\
\end{tabular}
\label{tab:2}
\end{table}

\begin{table}
\caption{The ground state energies and their degeneracies of the 
8-fold coordination for $w>0$ ($n=1,\,2,\,3,\ldots$)}
\begin{tabular}{lllll}
Class&Moment ($J$)&\multicolumn{2}{c}{Ground state}&First excited\\ 
     &      &energy&degeneracy&state energy \\\hline
1 & $3n$       & $-3w$         & 1 & $-w$       \\
2 & $3n\pm1/2$ & $-\sqrt6w$    & 2 & $0$        \\
3 & $3n\pm1$   & $-2w$         & 3 & $0$        \\
4 & $3n\pm3/2$ & $-\sqrt3w$    & 4 & $\sqrt3w$  \\
\end{tabular}
\label{tab:3}
\end{table}

\begin{table}
\caption{The ground state energies in magnetic field along 
$(1,1,1)$ direction and the low temperature magnetic susceptibilities 
of the 8-fold coordination classes}
\begin{tabular}{lll}
Class&Ground state energy&Susceptibility\\
\hline
1 & $-3w-J^2h^2/(6w)$ 			& $1/(3w)$  \\
2 & $-\sqrt6w-Jh/3-\sqrt6J^2h^2/(27w)$ 	& $\beta/9$ \\ 
3 & $-2w-Jh/2-13J^2h^2/(144w)$ 		& $\beta/6$ \\
4 & $-\sqrt3w-2Jh/3-\sqrt3J^2h^2/(54w)$ & $2\beta/9$\\
\end{tabular}
\label{tab:4}
\end{table}

\begin{table}
\caption{Distribution of $R^{3+}$ magnetic ions among the classes of 
the 6- and 8-fold coordinations.}
\begin{tabular}{llllll}
Class   & 1     & 2           & 3     & 4     & 5       \\ \hline
6-fold  & Pm Ho & Ce Nd Sm Gd &       &       & Pr Eu 	\\
	&	&  Dy Er Yb   &       &       & Tb Tm   \\ \hline 
8-fold  & Pr Eu & Ce Gd Yb    & Pm Ho & Nd Sm & 	\\
	& Tb Tm &	      &	      & Dy Er &		\\
\end{tabular}
\label{tab:5}
\end{table}


\begin{thebibliography}{99}
\bibitem{Haldane} F.~D.~M.~Haldane, {Phys. Rev. Lett. {\bf 50}, 1153 (1983)}; 
{Phys. Lett. A {\bf 93}, 464 (1983)}.
\bibitem{Haldane1} F.~D.~M.~Haldane, {Phys. Rev. Lett. {\bf 61}, 1029 (1988)}.
\bibitem{Loss} D.~Loss, D.~P.~DiVincenzo, and G.~Grinstein, 
{Phys. Rev. Lett. {\bf 69}, 3232 (1992)}
\bibitem{local1} V.~A.~Kalatsky, E.~M\"uller-Hartmann, V.~L.~Pokrovsky, and 
G.~S.~Uhrig, to be published in PRL.
\bibitem{Berry} M.~V.~Berry, {Proc. Roy. Soc. London, Ser. A 
{\bf 392},45 (1984)}
%\bibitem{Loss1} H.~B.~Braun and D.~Loss, 
%{\em Phys. Rev. B {\bf 53}, 3237 (1996)}.
\bibitem{CEF} M.~T.~Hutchings, in {\em Solid State Physics}, 
edited by F.~Zeitz and D.~Turnbull (Academic, New York, 1964), 
Vol. 16, p. 227. 
\end{thebibliography}
\end{document}